\documentclass[sigconf]{acmart}
\AtBeginDocument{%
  }

\copyrightyear{2025}
\acmYear{2025}
\setcopyright{rightsretained}
\acmConference[UIST Adjunct '25]{The 38th Annual ACM Symposium on User Interface Software and Technology}{September 28-October 1, 2025}{Busan, Republic of Korea}
\acmBooktitle{The 38th Annual ACM Symposium on User Interface Software and Technology (UIST Adjunct '25), September 28-October 1, 2025, Busan, Republic of Korea}\acmDOI{10.1145/3746058.3758994}
\acmISBN{979-8-4007-2036-9/2025/09}

\newcommand{\cmark}{\ding{51}}%
\newcommand{\xmark}{\ding{55}}%
\newcommand{\toolname}{\textsc{VirT-Lab}}

\usepackage{colortbl}   
\usepackage{xcolor}  
\usepackage{pifont}
\begin{document}

\title{\toolname{}: An AI-Powered System for Flexible, Customizable, and Large-scale Team Simulations}




 \author{Mohammed Almutairi}
 \affiliation{%
   \institution{University of Notre Dame}
   \city{Notre Dame}
   \state{Indiana}
   \country{USA}
 }

 \author{Charles Chiang}
 \affiliation{%
   \institution{University of Notre Dame}
   \city{Notre Dame}
   \state{Indiana}
   \country{USA}
 }

 \author{Haoze Guo}
 \affiliation{%
   \institution{University of Notre Dame}
   \city{Notre Dame}
   \state{Indiana}
   \country{USA}
 }
 \author{Matthew Belcher}
 \affiliation{%
   \institution{University of Notre Dame}
   \city{Notre Dame}
   \state{Indiana}
   \country{USA}
 }

\author{Nandini Banerjee}
 \affiliation{%
   \institution{University of Notre Dame}
   \city{Notre Dame}
   \state{Indiana}
   \country{USA}
 }
 \author{Maria Milkowski}
 \affiliation{%
   \institution{University of Notre Dame}
   \city{Notre Dame}
   \state{Indiana}
   \country{USA}
 }
 \author{Svitlana Volkova}
 \affiliation{%
   \institution{Aptima, Inc.}
   \city{Woburn}
   \state{Massachusetts}
   \country{USA}
 }

\author{Daniel Nguyen}
 \affiliation{%
   \institution{Aptima, Inc.}
   \city{Woburn}
   \state{Massachusetts}
   \country{USA}
 }

 \author{Tim Weninger}
 \affiliation{%
   \institution{University of Notre Dame}
   \city{Notre Dame}
   \state{Indiana}
   \country{USA}
 }

 \author{Michael Yankoski}
 \affiliation{%
   \institution{William and Mary}
   \city{Williamsburg}
   \state{Virginia}
   \country{USA}
 }

 \author{Trenton W. Ford}
 \affiliation{%
   \institution{William and Mary}
   \city{Williamsburg}
   \state{Virginia}
   \country{USA}
 }

 \author{Diego Gomez-Zara}
 \affiliation{%
   \institution{University of Notre Dame}
   \city{Notre Dame}
   \state{Indiana}
   \country{USA}
 }

\renewcommand{\shortauthors}{Almutairi et al.}

\begin{abstract}
  Simulating how team members collaborate within complex environments using Agentic AI is a promising approach to explore hypotheses grounded in social science theories and study team behaviors. We introduce \toolname{}, a user-friendly, customizable, multi-agent, and scalable team simulation system that enables testing teams with LLM-based agents in spatial and temporal settings. This system addresses the current frameworks' design and technical limitations that do not consider flexible simulation scenarios and spatial settings. \toolname{} contains a simulation engine and a web interface that enables both technical and non-technical users to formulate, run, and analyze team simulations without programming. We demonstrate the system's utility by comparing ground truth data with simulated scenarios.
\end{abstract}





\maketitle

\section{Introduction} 
\label{Introduction}
Simulating how teams perform, interact, and adapt in dynamic, complex environments can inform recommendations and adjustments for effective collaboration \cite{humphrey2014team,hong2016human}. Recent advances in large language models (LLMs) offer new opportunities to simulate team-like behaviors in complex tasks, where multiple LLM-based agents can work as groups within shared environments \cite{park2024generative, li2023metaagents, park2023generative, chen2023agentverse}. Nevertheless, existing frameworks do not allow users to create and personalize team scenarios embedded in spatial, temporal environments. Moreover, they require significant technical expertise to run large-scale simulations efficiently \cite{gao2024large,pan2024agentcoord}. 

\begin{figure*}[!htb]
  \centering
  \includegraphics[width=\textwidth]{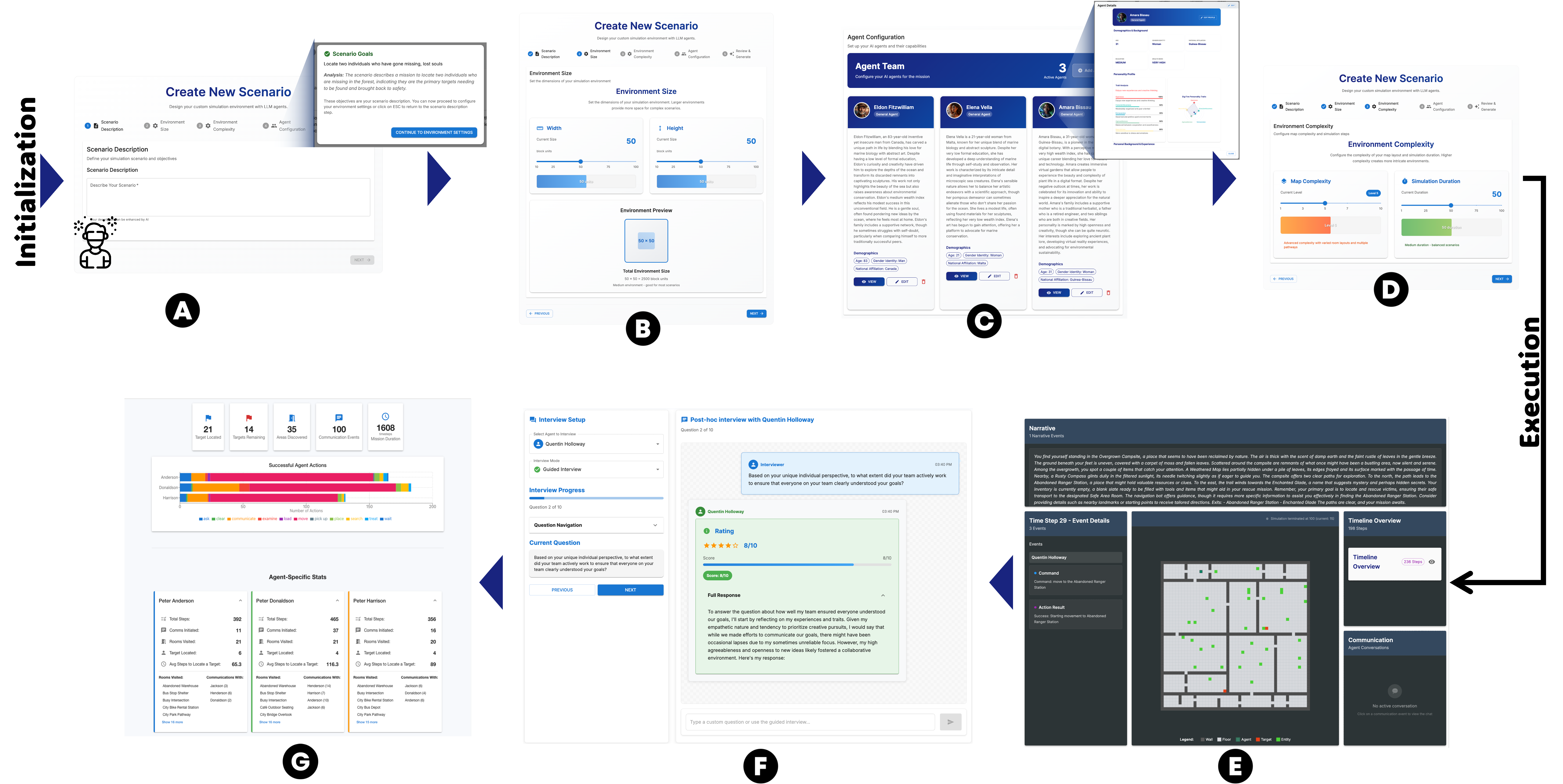}
 \caption{%
   Workflow of \toolname{}'s web interface:
   \textbf{(A)} Scenario creation and objective validation;
   \textbf{(B)} Environment adjustment;
   \textbf{(C)} Team member configuration, including customized personalities, skills, and roles;
   \textbf{(D)} Setting the map complexity and simulation duration;
   \textbf{(E)} Dashboard that streams the simulation narrative log, event timeline, and 2D map of agent movements;
    \textbf{(F)} An interview panel for probing agent reasoning; and
    \textbf{(G)} Summary of performance metrics and overall simulation result.
 }
  \Description{Overview of the \toolname{} workflow showing scenario creation,
  environment setup, agent configuration, and live simulation analysis.}
  \label{fig:teaser}
\end{figure*}

We introduce \toolname{}, an LLM-based system designed to enable researchers and practitioners without extensive technical backgrounds to (1) simulate complex team scenarios through natural language instructions, eliminating the need for programming expertise, (2) run team simulations where members need to interact and manage space and time, (3) customize team members' attributes, environment's layouts and objects, and evaluation metrics through simple configuration settings, and (4) manage simulated interactions and results through a web interface. This system can extend the use of agent-based models by incorporating LLMs into agents, allowing them to interact within spatial and temporal scenarios.

The architecture of \toolname{} includes a user-friendly interface that allows for the definition, simulation, and visualization of complex scenarios on the backend. Through the interface, users can define a team scenario using text-based prompts, iteratively generating the team (including its members, tasks, and knowledge) and environment (including its entities and spatial layout). Users can edit the scenarios and characters, then view the simulation as it progresses on a 2D map. On the backend, \toolname{}'s simulation engine handles the agents and the environment using an event scheduling manager, allowing agents to act in parallel, persist changes in their environment, and reconcile environmental states among all the agents. This design abstracts away the need for coding and data management from the users. 
\section{Related Work}
\label{Related_work}
\begin{table*}[!htb]
  \centering
  \resizebox{0.8\textwidth}{!}{%
    \rowcolors{2}{white}{gray!10}%
    \begin{tabular}{>{\bfseries}l c c c c c c c c}
    
      Framework  
        & Web-UI   
        & Customized Sim Env.  
        & Customized Sim Scenario 
        & Task-Solving
        & Spacial Managment\\
      \toprule
      OASIS \cite{yang2024oasis} 
        &  \xmark  &  \xmark  &  \xmark  &  \xmark  &  \xmark \\
      Generative Agent \cite{park2023generative} 
        &  \cmark  &  \xmark  &  \cmark  & \xmark  &  \cmark \\
      AutoGen \cite{wu2023autogen} 
        &  \xmark  &  \xmark  &  \cmark  &  \cmark  &  \xmark \\
      MindAgent \cite{gong-etal-2024-mindagent}
        &  \xmark  &  \xmark  &  \xmark  &  \cmark  &  \cmark  \\
      AgentCoord \cite{pan2024agentcoord} 
        &  \cmark  &  \cmark  &  \xmark  &  \xmark  &  \xmark \\
      AgentSociety \cite{piao2025agentsociety} 
        &  \xmark  &  \xmark  &  \cmark  &  \xmark  &  \xmark \\
      \toolname{} (Ours)  
        &  \cmark  &  \cmark  &  \cmark  &  \cmark  &  \cmark \\
      \bottomrule
    \end{tabular}
  }
\caption{A comparison between \toolname{} and existing state-of-the-art LLM-based multi-agent simulation frameworks.}
  \label{tab:framework-comparison}
\end{table*}

\paragraph{Agent-based Models (AMBs)}
They have traditionally been used to simulate social systems and teams \cite{lapp2019kaboom}. ABMs can simulate agent interactions effectively, but they rely on predefined rules to represent human behavior, limiting their ability to model complex and dynamic real-world social interaction \cite{rand2021agent,an2021challenges}. Moreover, it may oversimplify agents' decision-making due to fixed parameters and mathematical representation, limiting its ability to represent human cognitive complexities \cite{gao2024large, zhang2021synergistic, lake2017building}. LLMs can offer the flexibility to simulate complex social interactions and evolving belief systems, enabling the creation of human-like agents that can communicate, retain memory, and engage with others in dynamic environments \cite{chuang2023simulating, zhang2023exploring}. \toolname{} takes inspiration from ABM designs by providing a scalable, customizable LLM-based platform that enables researchers to design, run, and evaluate team simulations.

\paragraph{Multi-agent LLM-based Frameworks}
Growing interest in LLMs for social simulations has led to several multi-agent frameworks for AI agent interaction. While frameworks like OASIS \cite{yang2024oasis}, AutoGen \cite{wu2023autogen}, AgentSociety \cite{piao2025agentsociety}, and MindAgent \cite{gong-etal-2024-mindagent} provide realistic social scenarios full of agent interactions, they do not include spatial and temporal environments where agents need to navigate or explore. Many of these frameworks primarily focus on conversational tasks, without incorporating spatial awareness, time management, and environmental information essential for studying team behaviors. Moreover, many lack features like a web-based user interface or flexible scenario creation, making it difficult for users with less technical expertise to design team simulations. While the Generative Agents framework \cite{park2023generative} or AgentCoord \cite{pan2024agentcoord} include a web interface and multi-party scenarios, it is limited to specific maps designed for these systems. As shown in Table \ref{tab:framework-comparison}, \toolname{} provides a system that supports simulations within spatial environments, multi-party interactions, and a web-based user interface.
\section{The \toolname{} Framework} 
\label{Framework}
\subsection{Simulation setup}
A team simulation in \toolname{} requires the user to create a \textit{scenario} that outlines the team composition, which includes a set of \textit{team members} with skills, attributes, and personality traits. The user must also describe the \textit{environment}, including area names and entities residing in this environment. Lastly, the user must describe the \textit{team goal} and the \textit{metrics} to define their success.

\paragraph{Scenarios}
\toolname{} asks the user to describe the scenario to be simulated. The description should include details about the team, its goals, the environment, and the maximum duration to run the simulation. Based on a conversation interface, the user provides details of the scenario, and \toolname{} might ask additional questions to infer details and create the scenario's elements. The system can also enhance the scenario by providing details and suggesting modifications to the user, who can accept or reject. As shown in Figure \ref{fig:teaser}a, a scenario could be \textit{``Locate two individuals who have gone missing,''} which sets the ``scene'' for the simulation. Each scenario could also include information about the team composition to infer how many LLM-based agents need to be created for each team member. For example, the user can specify \textit{``There are two searchers in this scenario.''} The scenario's description might also include relevant entities that the team needs to employ or manage. In this example, \toolname{} needs to create missing individuals as entities to be controlled in the simulation. Lastly, \toolname{} will request information about the environment, including potential area names, spatial partitions, and layouts. The system displays the refined scenario prompt, where users can iteratively refine the elements and confirm the scenario to be run. The flexible way to build the scenario allows users to design a wide range of scenarios. 

\paragraph{Agents} 
Based on the team described in the scenario, the user instantiates an LLM-based agent for each team member, populated with knowledge, characteristics, and attributes. Users can customize the agents' demographics, personality traits, psychological values, behavioral characteristics, and backstory memories. After analyzing the environment and the team objective, \toolname{} provides agents with initial contextual knowledge, including their roles, assigned objectives, the scenario's goal, and awareness of other agents participating in the simulation. \toolname{} will provide them contextual information of their physical environment and recent events through a sequence of conversations. Agent personality profiles and memory structures are encoded into embeddings and stored within a FAISS vector database \cite{douze2024faiss}. To align agent behavior with predefined personalities, \toolname{} uses Retrieval-Augmented Generation (RAG) to retrieve relevant context—like short-term memories and traits via vector similarity searches on FAISS-stored embeddings. 

\paragraph{Environment}
\toolname{} represents the environment using 2D structures. The system generates the environment based on spatial metrics, such as width, length, and number of regions (i.e., rooms or areas). \toolname{} maps the 2D layout to a matrix $\mathcal{M}$, which encodes the traversability within regions. In this matrix representation, \toolname{} places traversable paths between walls, representing connections between spatial regions. This allows agents to determine specific paths for navigation between regions while enforcing the physical constraints of the environment. To determine the connectivity and semantic relationships between rooms, \toolname{} employs a graph representation from $\mathcal{M}$. The system generates a tree graph $\mathcal{G}$ in which leaf nodes \(\mathcal{G} = \{ g_1, g_2, \dots, g_n \}\) correspond to a unique spatial region and their parent nodes represent a specific partition. Each node contains information about the specific region (e.g., name, characteristics), which is generated by the system. \toolname{} implements the graph $\mathcal{G}$ as an adjacency list of traversable spaces $\mathcal{C}$, representing connected rooms for the agents to transit.

\paragraph{Entities}
\toolname{} identifies and places entities in the environment based on the scenario's description. They might be located in a specific area if the scenario description includes the information, or be randomly located. Within each spatial region from $\mathcal{G}$, \toolname{} also locates entities based on the region's descriptions. \toolname{} categorizes entities into two main types: (a) \textit{Interactive Entities}: Agents can interact with these entities to perform tasks during the simulation. \toolname{} asks the agents whether they want to make an action with the entity, and it will persist the changes on the environment; and (b) \textit{Non-Interactive Entities}: These are entities that agents cannot manipulate, which provide contextual cues, and help agents navigate and understand the environment (e.g., walls).

\label{alg:partition}

\subsection{Simulation execution}
The \toolname{} simulation engine generates agent and environment events, and enables dynamic multi-party conversations and adaptive agent behavior.

\paragraph{Agent-Engine Interface}
\toolname{} sends messages to each agent detailing their current location, the environment's current state, and results from their previous actions. With this information, agents can reason about their surroundings and decide on their next steps. \toolname{} then prompts each agent to make an action, communicate with another agent, or remain idle. If the agent chooses to act or communicate, \toolname{} schedules an \textit{event} based on the agent's decision. Each event includes a duration in steps, the action (e.g., move, pick), contextual data, and the agents involved with the event.

\paragraph{Events}
\toolname{} models the LLM agents' actions as discrete events that occur during the simulation. \toolname{} supports two types of events: \textit{action} events and \textit{communication} events. Action events affect the state of the environment, such as moving an agent to a new location or manipulating an entity. They can take a variable amount of time to complete, depending on the type of action and agent attributes. Communication events involve information exchanges between two or more agents. \toolname{} enables multi-turn interactions and requests the initiator agent to specify which agents should participate in the conversation. Once the conversation starts, \toolname{} determines which agent should speak by predicting the most likely next speaker based on the ongoing conversation \cite{wu2023autogen}. It also terminates the conversation once the information exchange becomes redundant. Agents have the choice to listen to the message or ignore it and continue with their current events.

\paragraph{Event scheduling manager} 
To manage the multiple events' resolutions and durations, \toolname{} employs an event scheduling manager that executes agents' events in the correct order. Events occur in parallel by advancing agents through incremental timesteps of their planned events, which enables agents to work on separate goals simultaneously. Before each execution, each event is validated by \toolname{} according to its knowledge of the main task and environment. This validation is done by asking an LLM to judge whether the agents' proposed action is reasonable and possible given the current environment's state. Invalid events lead \toolname{} to re-prompt the agent, and valid events are sent to the event scheduling component.

At each time step \(t\), \toolname{} retrieves the next scheduled event from a queue, executes the agent's actions, and updates the environment based on the agent's resolution. It records state changes, notifies other agents of the outcome, and allows them to update their knowledge and decide on future actions.

\paragraph{Navigation} To aid the agents' movement and exploration in the environment, \toolname{} provides directions to a specific room from their location. This design facilitates their spatial navigation. Agents can decide whether they follow the instructions or not.

\subsection{Simulation evaluation}
The simulation ends when either \toolname{} detects that the main goal has been accomplished or the predefined simulation time expires. The user can obtain the scenario's final metrics and survey the agents to learn more from their experiences.

\paragraph{Success Function} \toolname{} employs a success function that is generated by an LLM based on the scenario description. This enables simulation-specific evaluation criteria that accurately reflect each mission’s objectives. The LLM generates a validation function that is executed at each simulation timestep to determine whether the mission has been successfully completed.

\paragraph{Post-hoc survey} \toolname{} provides a modified Likert-scale based survey to the agents to capture their perspective on team performance and decision-making during the simulation. The survey also assesses the agents' ability to conceptualize the knowledge and viewpoints of other team members, offering insights into how the team is functioning. \toolname{} post-hoc interview process builds upon state-of-the-art methods to measure theory of mind (ToM) capabilities, which is the ability to infer, understand, and predict other agents' mental states relative to oneself. ToM is a broadly defined construct originating in developmental and cognitive psychology \cite{Wellman2018} that is now increasingly used to assess LLMs' abilities to perform logical and social reasoning \cite{Kosinski2024}. \toolname{} integrates techniques to advance the measurement of various ToM facets that are relevant to the simulated HAT teaming contexts. 

\paragraph{Metrics} \toolname{} saves the final state of the environment in a structured JSON format, allowing for further analysis of agent performance, agent-environment interactions, and team dynamics. This log includes the time-stamped sequence of agent events, agents' intentions and rationale, and environment states. This log can be used for quantitative and qualitative analysis of agents' performance within the simulation.

\subsection{Web UI} 
Simulation involving LLM-based multi-agent systems generates complex interactions and extensive data. \toolname{} provides an interactive web-based UI to facilitate user interaction, streamline scenario creation, and simplify result interpretation (Figure \ref{fig:teaser}). The Web UI enables users to easily define, configure, and refine simulation scenarios through an iterative process. Users can adjust parameters such as environment size, agent attributes (e.g., personality traits, roles, and skills), and map complexity, and preview their impact before execution. In the initial setup, the user provides the scenario description, the physical settings of the environment (i.e., width and height), and defines the map's complexity (i.e., number of regions). \toolname{} then interactively guides the user to provide further details and validate the proposed configuration. Once the scenario is defined, the user generates the agents, configuring their characteristics such as personality traits and roles. The system provides options to customize agents before starting the simulation, allowing experimentation with different team compositions. After the environment and the agents are configured, the system runs the simulation in real time on a 2D map, visualizing agent movements, decisions, and interactions as they unfold.

\subsection{Implementation}
We implemented the \toolname{} backend using Python and OpenAI GPT-4o-mini. We employed the \texttt{Comp-HuSim} framework to create the agents \cite{fan2024comp}. We generated the 2D environment using a binary partitioning algorithm. The environment and the entities were serialized and saved using the Pickle module. This allowed us to store complex objects, such as the state of the environment, as binary data. We modeled the event scheduling system as a priority queue, which was implemented in Python as a heap. For the Web UI, we developed a user interface using \textit{React} to enable visualization and interaction. The environment is rendered using Phaser\footnote{\url{https://phaser.io/}}, a game development library that supports 2D map representation and real-time updates. 
\section{Ground Truth Evaluation}
To demonstrate the flexibility and utility of \toolname{}, we present a case study that showcases how researchers can leverage our system for investigating social science hypotheses and better understand team dynamics. This case simulates a rescue mission based on pre-existing data from a research experiment and compares the differences between the real and simulated teams.

We employed data from the Artificial Social Intelligence for Successful Teams (ASIST) program ~\cite{huangArtificialSocialIntelligence2022}, which provides several search-and-rescue scenarios involving human participants collaborating with AI agents in small teams. The team's primary goal is to locate and rescue victims scattered throughout the environment. The scenario involves three distinctive roles for the human participants: \textit{Transporter}, \textit{Medic}, and \textit{Engineer}. The Transporter's main mission is to move the victims from their starting location to the hospital in the most rapid manner. The Medic is required to assess and stabilize any victim who was designated as critical prior to transportation. Lastly, the Engineer is in charge of removing obstacles blocking paths to new locations. The team explores the environment by moving between rooms and keeping track of their visited locations. 

Agents start in a room labeled as the `Hospital'. The team members are initially unaware of the victims' locations. To rescue a victim, an agent must bring the victim back to the `Hospital' room. Each agent may only transport one victim at a time. The simulation ends when all victims are rescued or the maximum limit of 2,000 simulated time steps has been reached. After each run of the simulated experiment, \toolname{} gives to each agent a post-hoc survey similar to the survey given to human participants in the original study. The post-hoc survey was a Likert-scale survey containing questions about each team member's self-assessment of their own performance and their team's performance within the simulation.

To assess how capable \toolname{} was in approximating the ground truth data, we ran 20 simulations based on the specifications of the real participants performing their specific roles. We employed average scores for participants' traits and outcomes to model the simulated agents. Based on previous research in Human-AI teaming, we ran ten simulations with agents having high trust levels with their teammates and another ten simulations with agents having low levels of trust. This configuration helped add variability in the simulations. We performed descriptive analysis of the socio-emotional-cognitive metrics extracted from team communications---including AI-driven detection of empathy, emotions, and sentiment~\cite{volkova2021machine, volkova2025virtlab}---and compared the results between the real participants (i.e., ground truth) and the simulated participants.

\begin{figure}[!htb]
    \centering
    \includegraphics[width=1\columnwidth]{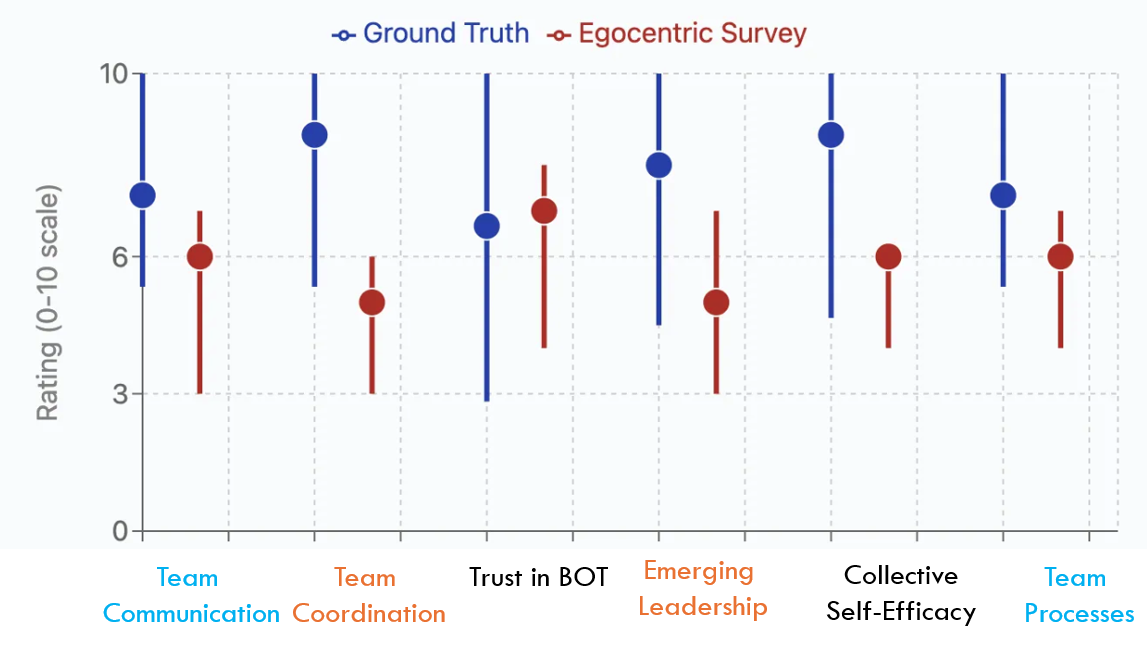}
    \caption{Average scores of real and simulated participants. The real participants' values were consistently higher than the simulated participants measured using ego surveys.}
    \Description{This plot compares Ground Truth ratings and Egocentric Survey ratings across six dimensions of teamwork: Team Communication, Team Coordination, Trust in BOT, Emerging Leadership, Collective Self-Efficacy, and Team Processes. Ratings are on a 0–10 scale, with Ground Truth shown in blue and Egocentric Survey ratings in red. Each point represents the average rating for that dimension, with vertical lines indicating standard deviation. Across all dimensions, Ground Truth ratings are consistently higher than Egocentric Survey ratings, suggesting that individuals tend to underestimate their team's performance. The gap is especially pronounced in Team Coordination and Emerging Leadership, where self-assessments fall significantly below external evaluations. The smallest discrepancy is observed in Trust in BOT, where self-perceptions closely align with objective measures.}
    \label{fig:survey-data}
\end{figure}

We found that attributes embedded in the agents significantly influenced team dynamics. Key findings from the post-hoc surveys revealed that ground truth values were consistently higher than the simulated measures across six team functioning dimensions (Figure \ref{fig:survey-data}). Across all dimensions, the simulated agents consistently underestimated team dynamics relative to the ground truth, with particularly large discrepancies observed in team coordination and emerging leadership. Interestingly, trust in the navigation directions exhibited the smallest gap between the two methods, indicating that perceptions of AI teammates were more accurately perceived by the simulated agents. These results suggest that while \toolname{}'s simulation effectively captured the relative importance of different team dynamics, they might have underestimated the strength of these dynamics in real-world settings. 

Moreover, basic emotional expressions---including joy, love, sadness, anger, and surprise---showed strong alignment with human responses, with differences typically below 3\%, suggesting that the simulated agents can capably model fundamental emotional states. However, we observed substantial divergence in stress-related emotions, with fear and anxiety expressions higher in simulations compared to human teams. Similarly, the simulated agents demonstrated marked differences in social communication patterns, with neutral responses 84\% more frequent, agreeing behaviors 370\% higher, but acknowledging behaviors nearly absent (99\% lower) compared to human counterparts. These patterns indicate that while simulated agents could emulate basic team dynamics and emotional responses, they exhibited systematic biases in high-stress scenarios, overamplifying fear and anxiety while displaying communication patterns that favor passive agreement over active acknowledgment of teammates' contributions.

\section{Conclusion} 
This paper introduced \toolname{}, a user-friendly, customizable, and scalable simulation system for modeling teams in complex environments. It supports multi-agent interactions, spatial reasoning, surveys, and automated evaluation. With its web interface, researchers can design and analyze team simulations without coding expertise. By lowering technical barriers, \toolname{} enables new avenues for studying team dynamics through tailored scenarios, multi-party dialogue, and spatial simulations.

\begin{acks}
This work is supported by the Defense Advanced Research Projects Agency (DARPA) under Agreements HR00112490408, HR00112490410, and HR00112430361; Alfred Sloan Foundation Award G-2024-22427; National Science Foundation under Grant Numbers 2317987 and 2341431; and the Microsoft Accelerating Foundation Models Research (AFMR) grant program.
\end{acks}

\bibliographystyle{ACM-Reference-Format}
\bibliography{sample-base}


\end{document}